# Effect of fixation target on the contrast sensitivity in the foveal and parafoveal area

Vicent Sanchis-Jurado,[a,*] Sophie Triantaphillidou,[b] Edward Fry,[b] Álvaro Pons[a]

[a]Universitat de València, Facultat de Física, Department d'Òptica Optometria i Ciències de la Visió, c/Dr Moliner 50, Burjassot, Spain, 46100

[b]University of Westminster, Faculty of Science and Technology, Computer Science, Computational Vision and Imaging Technology, 115 New Cavendish Street, London, United Kingdom, W1W 6UW

Corresponding author: Vicent Sanchis-Jurado, e-mail: visanju@gmail.com

## Abstract

Purpose

To determine the influence on the contrast sensitivity when the stimulus contains a fixation target in two retinal locations, foveal and parafoveal.

Methods

Four young adults with 0.0 logMar acuity participated in this study. The stimulus was based on vertical sinusoidal gratings masked by a circular (for foveal area) or a ring (for parafoveal area). To increase the luminance resolution of the display a bit-stealing technique was used. Four different sets of stimuli were generated, two for exploring the foveal sensitivity and two for the parafoveal area. The difference between the sets designed for the same area was the presence, or absence, of a fixation target (a white cross) in the centre of the stimulus. A modified staircase method was implemented.

Results

The results show a drop in the contrast sensitivity when the fixation target was present on the stimulus for frequencies smaller than 4 cycles per degree.

Conclusions

The presence of fixation targets diminishes the contrast sensitivity for low to mid frequencies over different concentric areas of the retina. This could be due to the fixational eye movements, different patterns of eye movements were found using an eyetracker. The relationship between the sensitivity in the foveal area and the parafoveal agrees with those reported by other authors using different designs confirming that the new stimulus design is suitable to measure the contrast sensitivity outside the foveal area.

# 1 Introduction

The contrast sensitivity function (CSF) is a well-known measure[1–6] for assessing the quality of vision and the visual performance[7] of refractive surgeries[8, 9], lenses (intraocular[10, 11] or contact lenses[12]) and other conditions[13] or illnesses[14]. From all the research that has been conducted around CSF models[2] several applications[13] have been developed and nowadays it is easy to find relevant commercial tests[15].

The effects of different parameters on the CSF, such as the type of grating, the size of the test, the temporal modulation, image stabilization[16], etc. are well established. Also, the effects of the eccentricity on the contrast sensitivity and visual acuity are quite well known [17–22]. Virsu[20], Johnston[18] and Thibos[19] chose a similar design for their studies. A sine-wave grating inside a circular patch, where the effect of the eccentricity explored was determined by the position of a fixation target. Kelly[17] chose a completely different design for their stimuli, i.e., radial gratings with increasing surface. His results showed a displacement of the band-pass shape towards lower spatial frequencies. However, Kelly's results are more difficult to compare because of the selection of the employed stimuli. Across the retina the sensitivity varies showing a different gradient depending on the direction[21, 22]. These asymmetries can be related to the asymmetry in the ganglion cells distribution across the retina[23].

The objective of the present study is to measure the central and peripheral contrast sensitivity using vertical sine-wave gratings masked by a circular and an annular mask. And to assess the differences between measuring the CSF under natural viewing conditions and when there is a fixation target and the only movements allowed are the fixational eye movements for the central and near periphery vision using an eyetracker device.

# 2 Methods

## 2.1 Participants

Four young adults, three men and one woman, participated in this pilot study. The mean age was 30 years old, with a range from 26 to 32. The inclusion criteria were healthy young adults with experience in psychophysical tests with a monocular visual acuity of 0.0 logMAR or better and a refractive error close to emmetropia (range between 0.5 dioptres of myopia to 0.5 of hyperopia and up to 0.5 dioptres of astigmatism) and an accommodation capacity of more than 2 dioptres. Exclusion criteria were a history of eye surgery, any kind of illnesses, to be under a drug treatment up to two weeks before the measurements and any problem on the tear film, cornea and/or pupil size. Subjects were informed by the experimenter about the nature and possible consequences of the experiment. The tenets of the Declaration of Helsinki were followed in this research. Informed consent was obtained from all individual participants included in the study.

## 2.2 Experimental setup

The experiment was conducted in a laboratory under total darkness. The 24 inch 8-bit per channel led computer display was placed 50 cm in front of a chinrest. This device was characterised using a photometer Tektronix J17 LumaColor (Tektronix, Beaverton, United States of America). The procedure to characterise the display and compensate for the non-linearity in luminance generation was as follows. First, five full-screen colour samples were generated per each colour channel with these digital values: 0, 64, 128, 192 and 256. The luminance of each sample was measured in the centre of the screen. The grey luminance was calculated by adding the three-colour luminances. A second-degree polynomial was fitted for each channel. Once knowing the luminance generated by each triplet of digital values a bit-stealing[24] own coded algorithm in Matlab generated six intermediate luminance levels between each pair of grey triplets by generating greyish colours. This way the luminance resolution can be increased by means of a pseudo-grey palette of 1785 values. For example, for the grey triplets (192, 192, 192), (193, 193, 193) and (194, 194, 194) the increasing in luminance is 0.4690 and 0.4718 $cd/m^2$, using the bit-stealing technique the contiguous triplets are (193, 193, 192), (193, 193, 193) and (193, 193, 194) and the increasing in luminance is 0.0395 and 0.0398 $cd/m^2$. The display was run by an Intel Core i5-3500 computer, with 16 gigabytes of RAM running Windows 10 64 bits. The volunteers had to adapt to the darkness inside the laboratory for 15 minutes before starting the test. Before starting the measurement the display was turned on for 30 minutes as a warm-up time.

## 2.3 Stimuli

Two stimuli were designed to assess the contrast sensitivity at different concentric retinal locations. One circle for the foveal region with a radius of 5.61 degrees, and a ring for the parafoveal with radii of 8.33 and 10 degrees. To test the influence of the gaze stability a white cross was placed in the centre of the test for both areas 36 pixels wide and 3 pixels of thickness, which is equivalent to 1.123 and 0.094 degrees of visual angle. Vertical sine-wave gratings of 0.5, 1, 2, 4, 6 and 8 cycles per degree were used. The edges of the stimuli were smoothed using a Gaussian function. The mean luminance of the grating was 40 cd/m$^2$. In order to increase the luminance resolution of the display the bit-stealing technique[24] was implemented as mentioned in the previous section. The smallest contrast step was 0,000495. We used the Michelson definition for contrast[15].

## 2.4 Psychophysical procedure

A modified staircase method[25] was used. The experiment consisted in showing first the reference image for 500 ms and after a test grey image (with the mean grating luminance) for another 500 ms. The task for the observer was to keep the eyes fixating in the centre of the stimulus and press the spacebar in case they perceived a difference between the test and the reference image. After three detections the contrast was diminished one step. In case the observers did not detect any difference between reference and test, they had to press any other key and then the reference image contrast was increased one step. After 4 reversals, the test stopped and the contrast threshold was calculated by averaging the four reversals. Sensitivity for the given spatial frequency was defined as 1/contrast threshold. The order between the different eccentricities, the presence of the fixation point and the frequencies were randomised, each combination was measured three times per observer and the results averaged.

## 3 Results

The averaged sensitivities for the four combinations of areas and presence of target are shown in figure. 1, table 1 contains the plotted values.

|  | With the fixation target | | | | Without the fixation target | | | |
| --- | --- | --- | --- | --- | --- | --- | --- | --- |
|  | Foveal | | Parafoveal | | Foveal | | Parafoveal | |
| Frequency (cpd) | Mean | Standard error | Mean | Standard error | Mean | Standard error | Mean | Standard error |
| 0.5 | 138 | 10.39 | 82 | 11.55 | 168 | 20.79 | 104 | 12.12 |
| 1 | 201 | 18.48 | 106 | 21.94 | 282 | 38.68 | 137 | 16.17 |
| 2 | 281 | 19.63 | 102 | 17.32 | 304 | 24.25 | 129 | 24.25 |
| 4 | 195 | 19.05 | 69 | 12.70 | 228 | 55.43 | 72 | 7.51 |
| 6 | 138 | 13.86 | 34 | 5.20 | 134 | 13.86 | 35 | 4.62 |
| 8 | 100 | 13.28 | 15 | 1.73 | 99 | 16.17 | 17 | 1.73 |

**Table 1** Average sensitivity for the different combinations

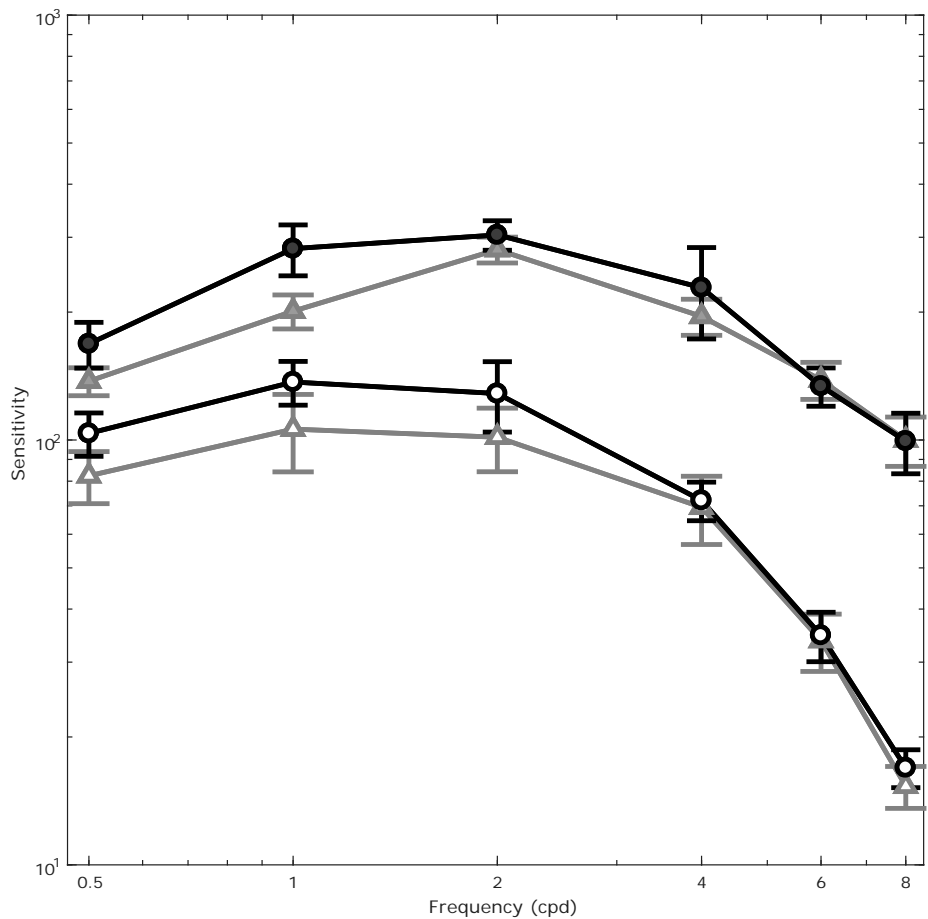

**Figure 1** Mean contrast sensitivity. Solid markers for the foveal region, open markers for the parafoveal. Circles for the test without the fixation target, triangles for the test with the fixation target

# 4 Discussion

Results show a reduction in the sensitivity when testing the same retinal area depending only on the presence of a fixation target, the sensitivity being higher when with no fixation target. This effect is more noticeable in low spatial frequencies as can be seen on fig. 2. The loss in sensitivity is noticed both in the foveal area and in the parafoveal area.

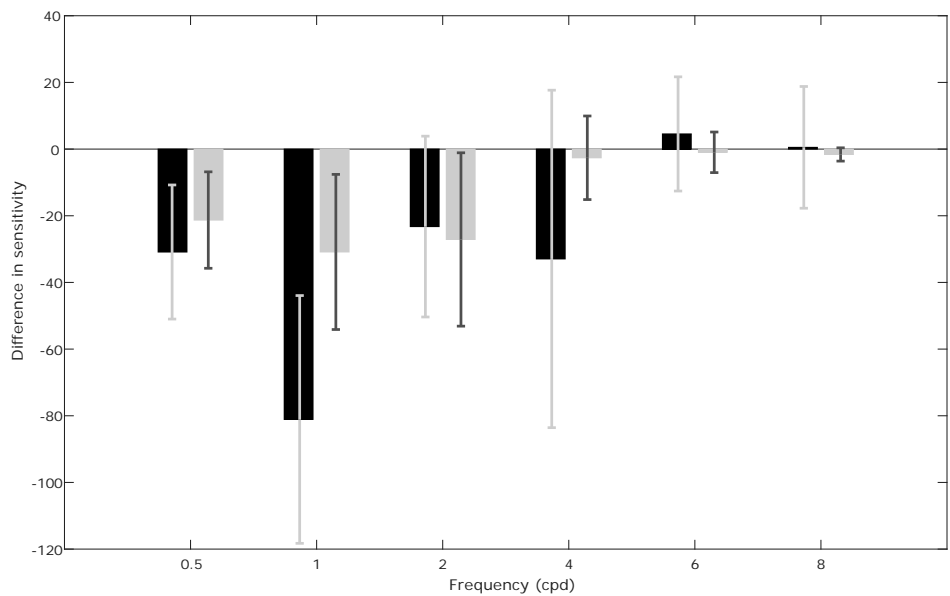

**Figure 2** Mean difference in sensitivity between 'with fixation target' and 'without fixation target' for the two retinal locations

As far as we know, this is the first study in which the peripheral sensitivity has been measured using a ring masking a sine-wave grating instead of a circular patch. It is known that the sensitivity in the periphery of the retina is not homogeneous [21, 22], but due to the nature of the fixational eye movements[26, 27] we considered this design the best option to prevent any bias imposed by the position of the stimulus. At the same time the annular configuration permits to change the viewing conditions between monocular and binocular without any change on the stimulus.

The results obtained without the fixation target agree with those reported by other researchers [17–20]. The sensitivity diminishes when the eccentricity of the stimulus is increased and the maximum of the CSF is displaced towards lower frequencies[19].

We found differences in the sensitivity between "with the fixation target" and "without the fixation target" stimuli for both areas. The differences are more noticeable for the lower frequencies, especially for 1 cycles per degree (cpd) in foveal vision. The effect of the fixation target over the stimulus spectrum is small. Taking two different stimulus, 1 and 6 cpd, and analysing their spectra, little differences can be seen (fig. 3); this would not justify any significant change in the contrast sensitivity.

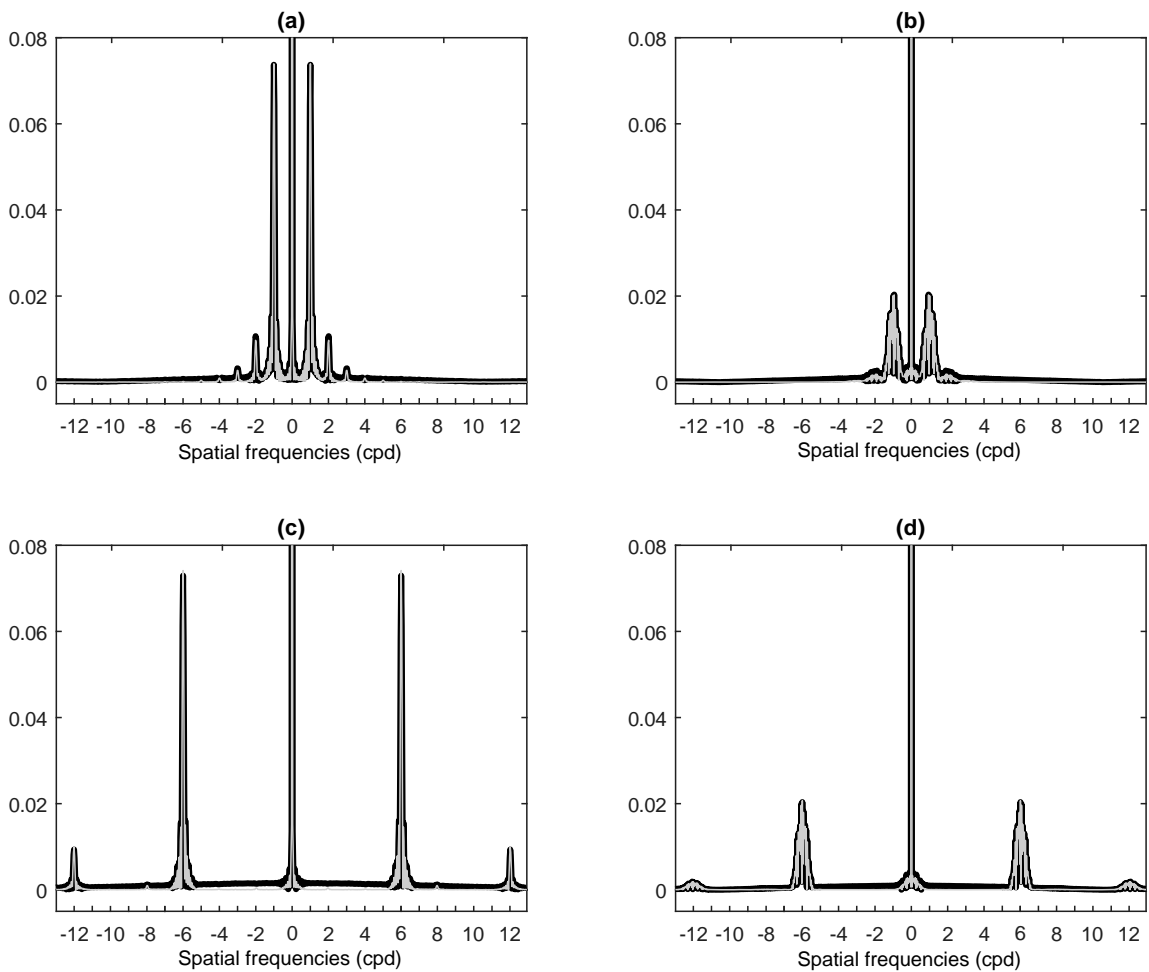

**Figure 3** Spectrum of the stimuli. Black for 'with the fixation target', grey for 'without the fixation target'. (a) foveal vision, 1 cpd, (b) parafoveal vision 1 cpd, (c) foveal vision 6 cpd, (d) parafoveal vision 6 cpd

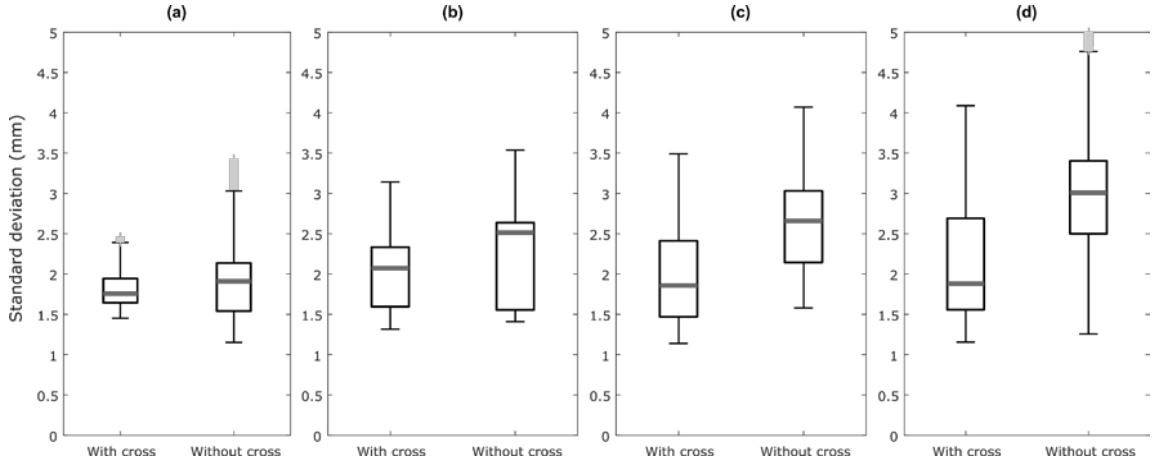

**Figure 4** Boxplot of the standard deviations for a temporal window of 4 seconds. (a) for foveal vision, 1 cpd; (b) for parafoveal vision, 1 cpd; (c) for foveal vision, 6 cpd; (d) for parafoveal vision, 6 cpd

The other factor that should be considered is the gaze stability due to the presence of the fixation target. The fixation target acts as a stimulus for the gaze stability and for the accommodation and convergence as well[28]. Using an infrared eyetracker (High-Speed Video Eye Tracker Toolbox, Cambridge Research Systems Ltd, Rochester, Kent, United Kingdom) we checked the gaze position of the right eye of one of the volunteers, while doing the test for two different frequencies. These chosen frequencies were 1 and 6 cpd, because they are representative of the two different behaviours described. The eyetracker captures the information about the pupil centre and the pupil size with a temporal frequency of 250Hz. To analyse all this information we chose a temporal window of 4 seconds (1000 values). This window was placed in the beginning of the sequence and moved towards the end moving step by step. The standard deviation was calculated for all the positions of the window and the results are represented in the figure 4.

For 1 cpd and foveal vision, in the horizontal direction the dispersion of the deviations is lower for the stimulus with the cross in comparison to the stimulus without the cross. A small height in the boxplots of fig. 4 means a small variation of the deviations over time, and this can be interpreted as a more stable fixation over time.

These results agree with the experiment done by Gilbert that compare the change in the threshold sensitivity between stabilized and non-stabilized gratings[16]. They found a significant reduction in sensitivity when the test grating was stabilized. This reduction had more effect in frequencies lower than 0.5 cycles per degree. In our study we are restricting the eye movements (except from the fixational eye movements) using the fixation cross. Gilbert's goal was to measure the effects of stabilization of the retinal image, while ours is to stabilize the fixation.

The influence of the fixation point on the sensitivity have been studied by Summers[29]. Their results are in disagreement with the ones that we report in the present study. Summers reports an increase in the contrast threshold (reduction in the sensitivity) in 4 and 8 cpd; and a little or almost no effect in 1 and 2 cpd, in our study, the effects are the contrary. We attribute the difference in the results that the differences between the stimuli in the two studies. Summers used a black square and we use a white hair cross of 3 pixels of thickness, they reported a small reduction in the masking effect when changing the colour of the fixation point to white. They found a relation between the size of the square, the size of the patch (or the number of visible cycles) and the masking effect. In fact, in their experiment 4, they compared the effects of the fixation point for 4 cpd between a patch of 0.75 degrees and another of 3 degrees. The results showed an increase in the threshold for the small patch but no effect for the big patch. In comparison, we are using a patch with a surface almost five times bigger.

In conclusion, the use of a fixation target has effects over the sensitivity values for a wide range of low to mid frequencies over different concentric areas of the retina. This reduction can be related to the fixation stability. As Pons[30] demonstrated, fixational eye movements act as a low-pass filter degrading the quality of the retinal image. In this pilot study we found different patterns of eye movements for different frequencies of the test that can be related to relevant loss in sensitivity. The use of a ring mask to measure the peripheral sensitivity provides results compatible with those reported by other studies. We consider necessary to further test the effects on contrast sensitivity of other designs for the fixation stimulus and to improve the integration of the eye tracking techniques as fixational eye movements play a role in vision but in general are not considered in vision research.

# Acknowledgments

This study was funded by Generalitat Valenciana (VALi+d, grant number ACIF/2014/142) and MINECO grants: TIN2013-50520-EXP and TIN2015-71130-REDT.

# Disclosure of potential conflicts of interest

The authors declare that they have no conflict of interest